\documentclass[traditabstract]{aa}

\usepackage{graphicx}

\newcommand{\kms}{~km~s$^{-1}$}
\newcommand{\degree}{\mbox{$^{\circ}$}}			% degree
\newcommand{\KS}{\mbox{$K_{\rm s}$}}
\newcommand{\LAM}{\mbox{$\tilde{\Lambda}_\odot$\,}}
\newcommand{\BET}{\mbox{$\tilde{B}_\odot$\,}}

\begin{document}
\title{A catalogue of oxygen-rich pulsating giants in the galactic halo and the Sagittarius stream
\thanks{}}
\author{N.~Mauron\inst{1}, L.~Maurin\inst{2}, T.R.~Kendall\inst{3}}
\offprints{N.~Mauron}

\institute{Universit\'{e} de Montpellier, Laboratoire Univers et Particules de Montpellier CNRS/UM,    
 Place Bataillon, 34095 Montpellier (France). 
  \email{nicolas.mauron@umontpellier.fr}
\and  Observatoire des Ifs, 6 impasse des Ifs, 84000 Avignon (France)
\and  Northampton NN1 4RG (United Kingdom)
}

\date{Received by A\&A: 16 August 2018 / Accepted 27 March 2019}

\abstract { In order to construct a catalogue of oxygen-rich (noted M)  asymptotic giant branch (AGB) stars in the  halo, complementing those of carbon-rich (C) stars, previous lists of Miras and SRa semiregulars  located in the northern hemisphere  are merged and cleaned of various defects. After putting aside known  C stars, characteristics like colours and periods indicate that most of the remaining objects are M stars. Distances are obtained through the period-luminosity relation. By considering their position in the sky,  stars lying at $|Z| > 5$~kpc are confirmed to be in majority in the Sgr tidal arms. The M stars are more numerous than C ones. Our distance scale is supported by two cool variables located in the Pal~4 globular cluster. Along the Sgr arms, there is reasonable agreement on distances of our objects with recent RR Lyrae distances. A few stars may be as distant as 150~kpc, with possibly four at the trailing arm apocenter, and two in the A16 substructure, angularly close to two C stars. 
Ninety radial velocities are collected from $Gaia$ and other sources. A catalogue with 417 M pulsating AGB stars is provided. It contains  $\sim$ 260 stars in the halo with  $|Z| > 5$~kpc. Their $K_{\rm s}$ magnitudes range from 8 up to 13. For comparison, the catalogue also provides $\sim 150$ stars in the disc having $5$~$< K_{\rm s} <$~$8$. {\it The catalog is available on request  (NM) before being on CDS.}} 

\keywords{ stars: AGB -- Galaxy: halo -- Galaxy: stellar content } 
\titlerunning{Oxygen-rich AGB stars in the galactic halo and the Sgr arms}
\authorrunning{N.~Mauron et al.}
\maketitle

% ============================================================================

\section{Introduction} 

Stars evolving on the asymptotic giant branch (AGB) can be either oxygen-rich (M), or carbon-rich (C). The general properties of AGB stars are reasonably understood and summarized in the book of Habing \& Olofsson (2004). Much previous research concerning   Mira variables  enabled a greatly improved knowledge of the main components of the Galaxy (bulge, disc, bar, etc). For example, by considering  Mira stars close to the Sun, their locations and their kinematics, Feast \& Whitelock (2000) could detect the extension of the galactic bulge bar in the solar vicinity and beyond.  In a galaxy, the presence of AGB C stars is generally interpreted as indicating that an intermediate-age population exists. In contrast,  the M AGB stars may be much older since one finds them in  globular clusters. 
The goal of this work is to achieve a catalogue of the M AGB population of the galactic halo, and we restrict this research to the Mira-type and SRa-type variable stars. These stars are periodic, and, in this paper, we name them MSRa stars. Also,  although this paper is devoted to M stars,  a detailed comparison with  C stars is  done at several points throughout the paper to underscore similarities and differences.

 In comparison with the galactic  disc, the understanding of AGB stars located in the halo has to be improved. A great deal of research has been carried out on C stars because their discovery is made easier owing to their very red colour, or  to their peculiar spectra; see Huxor \& Grebel (2015; hereafter HG2015) for a synthesis. However, in comparison to   C stars, relatively little attention was devoted to find specifically  M stars. The fact that one can find them in globular clusters suggests that there may be such a population  in the halo, but, to our knowledge, a systematic census is missing.

One of the  difficulties in finding M AGB stars at high galactic latitude is  to separate them from the large amount of dwarfs in the disc. Infrared properties and/or variability may be thought as useful  methods. 
One of the early studies of these M stars was based on the availability of the infrared IRAS data. Indeed, a complete survey of M Miras with short or intermediate period and with $|b| > 30^{\circ}$ is due to Jura \& Kleinman (1992), but  distances  to the galactic plane ($Z$) are in majority less than 3\,kpc. More precisely, of a total of 318 objects, only 13 have $|Z| >$~3\,kpc. Another interesting work is the one of Whitelock et al.\ (1994, 1995). To investigate cases of very high-mass loss AGB stars,  they choose them in  the South Galactic Cap at $b < -30^\circ$ and select those having IRAS fluxes and colours. Because fluxes at 12$\mu$m and 25$\mu$m are needed, these sources are not very distant: their time-averaged  magnitude is typically $K <$ 7, with $|Z| < 4$\,kpc. Finally, Gigoyan \& Mickaelian (2012) published a list of M stars (dwarfs or giants or AGB stars) found by eye scanning the Byurakan objective-prism plates. We have cross-matched their sample  with the General Catalogue of Variable Stars (Samus et al.\ 2017) and  85 Miras are identified.  However, their $K_{\rm s}$ magnitudes are less than 7 because the Byurakan plates are not deep.

This situation  has changed with large-scale experiments looking for variable objects over all the sky. A first catalogue of periodic giants  based on the LINEAR experiment was published by Palaversa et al.\ (2013). It covers 10000 square degrees and reaches $r_{\rm SDSS}$=18.   A second catalogue is the one by Drake et al.\ (2009; 2014), from the Catalina experiment, covering over 20000 square degrees down to $V \sim $\,19. In addition to yielding a number of new results on  variable objects in general, these databases dramatically increase the number of  known MSRa out of the Milky Way plane. From their Catalina sample of $\sim$ 500 MSRa, Drake et al.\ (2014)   show that some of these trace the Sgr stream.  We decided to reanalyse their data for several reasons. First, we intend to build a catalogue of M stars in the halo, as a complement to the HG2015 compilation of C stars.  Secondly, in order to determine distances and since periods are provided, we shall use the $K$-band period-luminosity relation instead of an  absolute magnitude $M_{\rm v} \sim -3$ adopted by Drake et al.~(2014). A third reason is that the Sgr arms have been mapped recently in detail with RR Lyr variables from the PanSTARRS data, providing accurate distances to which those of MSRa stars can be compared (Sesar et al.\ 2017; Hernitschek et al.\ 2017). Finally, we wish to investigate the possible existence of very distant stars, at more than $\sim$\,70\,kpc from the Sun. 
 
 In Sect.~2, we construct a preliminary list of MSRa stars  from catalogues mentioned above, pay attention to a variety of flaws and obtain a sample of 417 stars. In Sect.~3, we focus on their properties, in particular those independant of the distances: we examine their colours, (peak-to-peak) amplitudes, periods, apparent positions
 and how to best attribute a pulsation mode through consideration of photometric colours. In Sect.~4, results are described and discussed. Distances and their uncertainties are derived. It is found that the number of M stars in the halo from our sample is richer by a factor of $\sim$ 3 than the  presently known sample of halo C  stars. A few remarkable stars are discussed.
Our catalogue of  M stars is presented, and includes 90 radial velocities. 
Conclusions are finally given in Sect.~5.

\section{Building the sample}

To build a preliminary list of MSRas, we  consider first  the catalogue of Drake et al.\ (2014). Note that this catalogue is based on the Data Release 1 (DR1) of Catalina, and that supplementary monitoring data are provided in the DR2, which is usable on the Catalina web site. We add the MSRa stars contained in the catalogue of Palaversa et al.\ (2013). We also add  the C stars, periodic or irregular, compiled by HG2015, as well as other C stars that are warmer, but may have been missed as pulsators in previous studies. Finally, we examined stars that had been found as contaminants in our survey for C stars. After checking  their light curves on the Catalina DR2 database, those which were periodic were included. 

Because the LINEAR or Catalina coordinates  have an accuracy of $\sim$3\arcsec, we cross-correlated our list with the 2MASS catalogue with a search radius of 5\arcsec. Objects lacking a 2MASS counterpart were rejected. In rare cases, it may happen that more than one 2MASS source are found. Generally, there is one bright source and the others are much fainter. Then consideration of magnitudes allows to identify the true counterpart. However, when the choice is not obvious, the object was dropped. 

This first 2MASS cross-matching procedure leads to a list of 894 entries. Further cleaning  is necessary for many reasons. Numerous objects are common to Drake's and Palaversa's catalogues, and there are 112 duplicates in our preliminary list.  These duplicates are however interesting because they can inform about differences in amplitudes and periods provided by the two databases. There are also objects presented in the LINEAR or Catalina catalogues as being periodic variables, but a systematic examination  of Catalina light curves  shows them to be of poor quality, or almost constant, or  not clearly periodic. These objects were rejected, as well as those that are saturated. 
We also reject objects with  $\KS < 5$ in the 2MASS catalogue, because they are too close to us, but we  keep many  disc stars with typically \KS $\sim $ 5-7, in order to be able to compare halo/disc  properties.

 Occasionally, a periodic light curve displays a significant amplitude variation ($\sim$ 30\%). Although this may happen for AGB stars (see below),  another possibility is that the star is a still an unknown, slowly rotating, spotted giant. In the galactic bulge, the overwhelming majority of these variables have periods less than 100 days (Drake 2006). To our knowledge, no similar information exists for the thick disc or halo giant populations. This limit of 100 days is smaller than the periods of most of the stars studied in this work. We chose in these occasional cases not to be  too conservative at this stage, and we keep these rare cases in our list.  An additional reason for doing so is that supplementary observations could resolve this ambiguity. For example, spectra of red, variable members of  47 Tuc ([Fe/H] $\sim$ $-1.5$) display strong, variable Balmer emission lines  that are explained with  pulsation shocks (van Loon et al.\ 2007; their fig.~21). 

We also interrogated  Simbad at CDS and found that two quasars are present in our list: one is at $\alpha$=26.13977 $\delta$=27.08419, and the other at $\alpha$=116.99665 $\delta$=20.87373, coordinates J2000 in degrees. As a consequence, we systematically looked for matches in the QSO-AGN catalogue of V\'{e}ron-Cetty \& V\'{e}ron (2010;  168\,940 entries), and the Sloan DR7 quasar catalogue (Schneider et al.\ 2010; 105\,783 entries), but no supplementary cases were found. 

Finally, a more intricate cleaning necessity appeared when we focused our attention on very distant stars, possibly located at more than $\sim$\,100\,kpc. We discovered  that many of these faint sources ($K_{\rm s}$\,\,$\sim$\,12-14) with apparent periodic signals were in fact objects polluted by bright Miras located  at less than 2$'$ away and with identical periods.  A thorough search  within 5$'$ for all our sources resulted into identifying  nine cases, given in Table~1. 

To conclude,  this cleaning procedure  being achieved, 417 objects remain that are not known as being carbon rich. This is what we call our sample in the following.

\begin{table}[!t]
\caption[]{False very faint variables. Their (J2000) coordinates are in given degrees. 
$K_{\rm s}$ is the faint star magnitude from 2MASS. $P_1$ (in days) is its period from Catalina DR2. 
The polluting star name is indicated. $P_2$ is its period from the General Catalogue of Variable Stars. 
 $\rho$ is the angular separation in arcminutes.}
\begin{tabular}{rrrrrll}
\noalign{\smallskip}
\hline
\hline
\noalign{\smallskip}
 R.A.~~ & Dec.~~ & $K_{\rm s}$~~ & $P1$ & Mira & $P2$~~ & $\rho$\\
\noalign{\smallskip}
\hline
\noalign{\smallskip}

 57.8024 & 33.0446 & 16.62 & 419 & RX Per   & 422  & 0.95\\
58.2531  & 25.5515 & 13.87 & 225 & SX Tau   & 225  & 0.86\\
70.5752  &  6.8905 & 15.60 & 413 & BZ Tau   & 400  & 1.11\\
119.2810 & 21.3302 & 14.77 & 338 & XY Gem   & 340  & 0.78\\
257.7185 & 27.0710 & 13.65 & 299 & RT Her   & 298  & 1.02\\ 
259.7585 & 8.5025  & 14.39 & 266 & V477 Oph & 267  & 0.67\\
317.8590 & 13.3561 & 15.75 & 277 & AN Peg   & 280  & 1.25\\
326.2447 & 12.7085 & 11.91 & 314 & TU Peg   & 321  & 1.54\\
333.0724 & 14.5785 & 14.72 & 400 & RS Peg   & 415  & 1.53\\

\noalign{\smallskip}
\hline
\noalign{\smallskip}
\end{tabular}
\end{table}

% ==========================================================

\section{Analysis}
Since this paper deals with the halo MSRa population, including the Sgr arms, and because we do not have a complete  kinematic information for the majority of sources in our sample, we need to consider a quantitative limit separating halo stars and those of the Sgr arms on one side, and a complementary group of objects  from the disc, on the other side. After some tries, we found  that a separation with the height $Z$ (in kpc) from the galactic plane works reasonably well, and  that $|Z| = 5$\,kpc  is a good compromise. The quantity $Z$  is derived by assuming all objects to obey the \KS-band period-luminosity relation with Wesenheit indices of
Soszy\'{n}ski et al.\ (2007). 
A more detailed discussion on the determination of distances is done in Sect~4.1. Taking $Z$ smaller than 5\,kpc (i.e. 2-3~kpc) retains too many disc sources which pollute the halo subsample. Adopting a height of $\sim$~10\,kpc is too large because, at least from the model of Law \& Majewski (2010), the Sgr arms distance can be as close as  10~kpc from the Sun (see also Fig.~15 of HG2015).  In the following, results concerning the colours, amplitudes, periods, and location in the sky are presented. These characteristics are the best determined because they do not involve distances.

 \subsection{Colours, amplitude and periods}

\begin{figure}
\begin{center}
\includegraphics*[width=7cm, angle=-90]{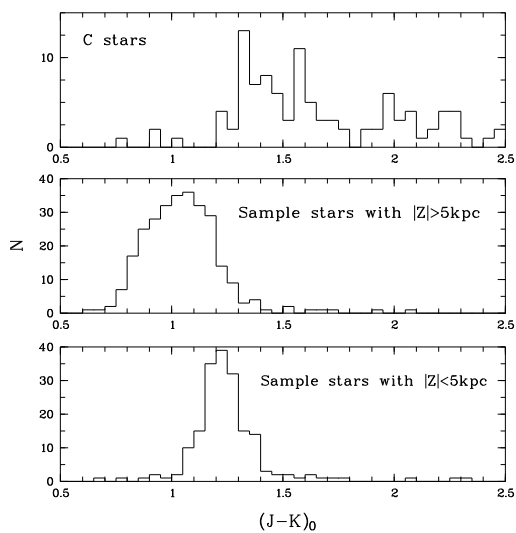}
\caption[] { Histogram of  $(J-K)_{\rm 0}$  of long period variables. Top panel: halo C  stars. Middle panel: stars in our sample for which the height $|Z|$ from the galactic plane is larger than 5~kpc. Lower panel: as middle panel, but $|Z| <$\,5\,kpc. In none of the panels do we include irregular variables, with no well-defined periodicity (they are ignored in this paper).}
\end{center}
\end{figure}

In Fig.\,1, we show three histograms of the $(J-\KS)_{0}$ colour index. Correction for interstellar extinction is achieved with $A_J = 0.87$\,$ E(B-V)$ and $A_{\rm Ks} = 0.35$\,$ E(B-V)$, from Cardelli et al.\ (1989), with $E(B-V)$ from Schlegel et al.\ (1998). The upper panel panel displays C stars. Stars of our sample  with $|Z| > 5$\,kpc and $|Z| < 5$\,kpc are shown in the middle and bottom panels, respectively.  It can be seen that  C stars are in very large majority redder than $(J-\KS)_{0} \approx $ 1.3, while an overwhelming proportion of our stars at $|Z| > 5$\,kpc are bluer. This strongly suggests that the latters are mostly oxygen-rich. The histogram ordinates are the number of objects per bin, and shows that the bluer sample  is considerably richer than the C star sample, by a factor $\sim$\,3.  The $(J-K_{\rm s})_0$ colours in the lowest panel suggest that the stars of our sample with $|Z| < 5$\,kpc are in majority M stars, although  a few C stars may be present. 

\begin{figure}
\begin{center}
\includegraphics*[width=7cm, angle=-90]{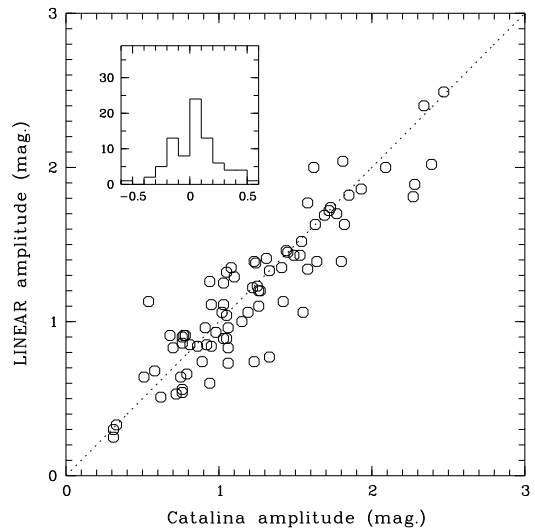}
\caption[] {Catalina versus LINEAR $V$-band amplitudes for common MSRa variables. The dotted line is a 1:1 line. The inset shows the histogram of residuals. The 1$\sigma$ scatter is 0.21 mag.}
\end{center}
\end{figure}

Pulsation amplitudes deserve  attention because they are involved in the  uncertainty on distances. This is because distance determination is based on the use of $\KS$, but the phase of this $\KS$ measurement is ignored. Therefore we attempt to estimate the range of variation in $\KS$  by scaling optical amplitudes to the near-infrared. The LINEAR survey took place between 1998 and 2009, while the Catalina survey DR1 covers the period 2005-2011. Thus, an overlap of four years exists. Consequently, we expect that, in a first approximation, amplitudes given by both experiments agree. Note that Catalina amplitudes are derived from Fourier fits, while LINEAR amplitudes are the range of values between 5\% and 95\% in the signal distribution. Figure 2 displays one amplitude $versus$ the other for common objects of our initial list. The agreement seems very reasonable, with $\sigma = 0.21$\,mag. This scatter means  that some amplitudes may change over years by at least this amount.

 In Fig.1, we saw that the halo sample stars (with $|Z| > $\,5~kpc) are generally bluer than the halo C stars, although there are exceptions. Here, we investigate whether a period-amplitude diagram would bring some complementary information.  Figure 3, panel a, shows that these halo stars comprise a main population with $P = 150$-$175$ days, and a small population with low amplitude and longer periods. In contrast,  panel b shows that the periods of C stars are more scattered over a 150-300 days interval, and do not include the  large-period low-amplitude family. Therefore, a periodic AGB halo star with period $P < 170$ days is  more probably of M type than of C type. This result is supported by an on-going program of M/C classification of several hundreds of Catalina MSRa in both hemispheres (Gigoyan, priv. comm.). This program shows that, at least for $V \leq 16$, only $\sim 3$\%  are newly discovered C stars.

\begin{figure}
\begin{center}
\includegraphics*[width=6cm, angle=-90]{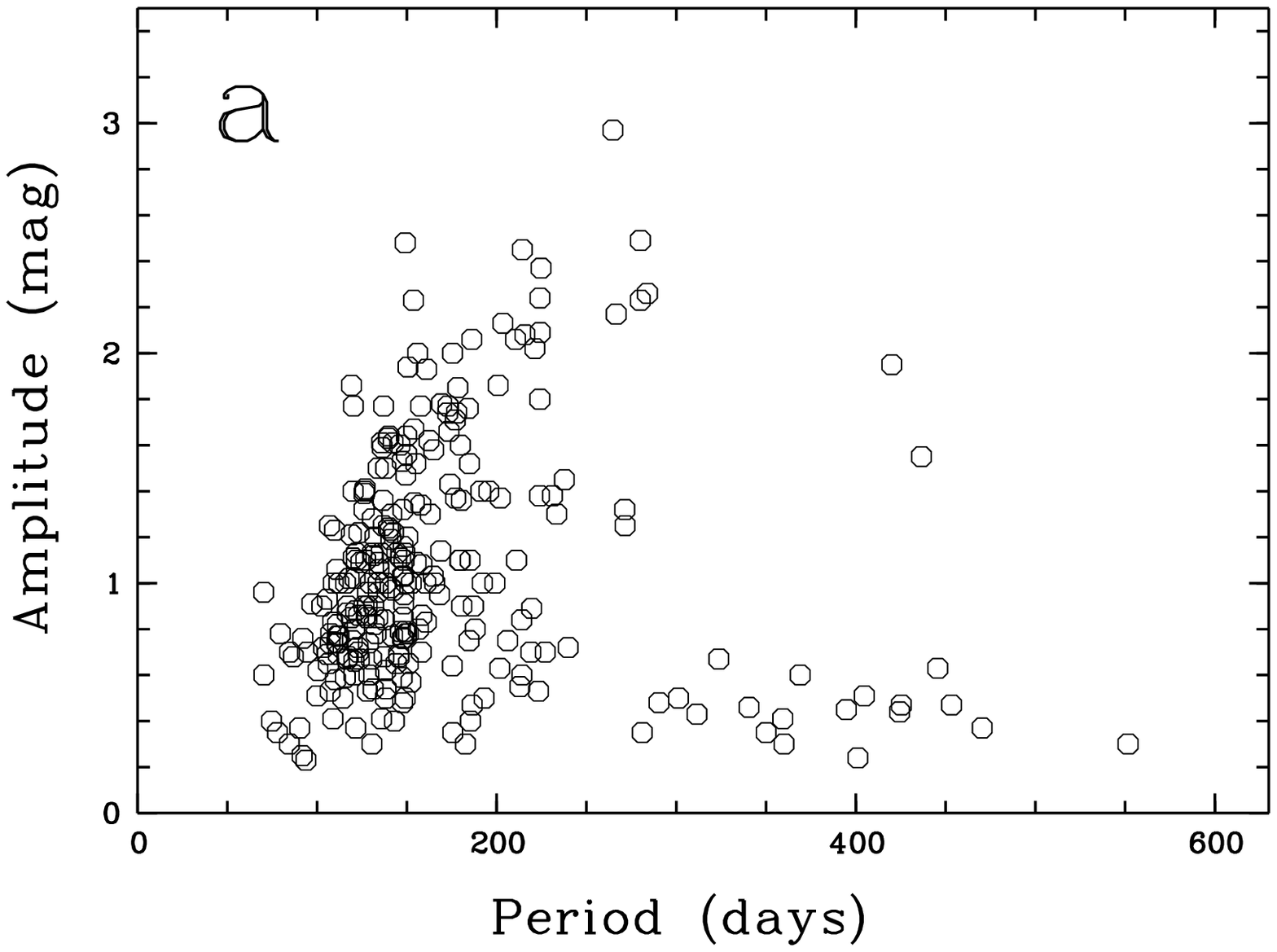}
\includegraphics*[width=6cm, angle=-90]{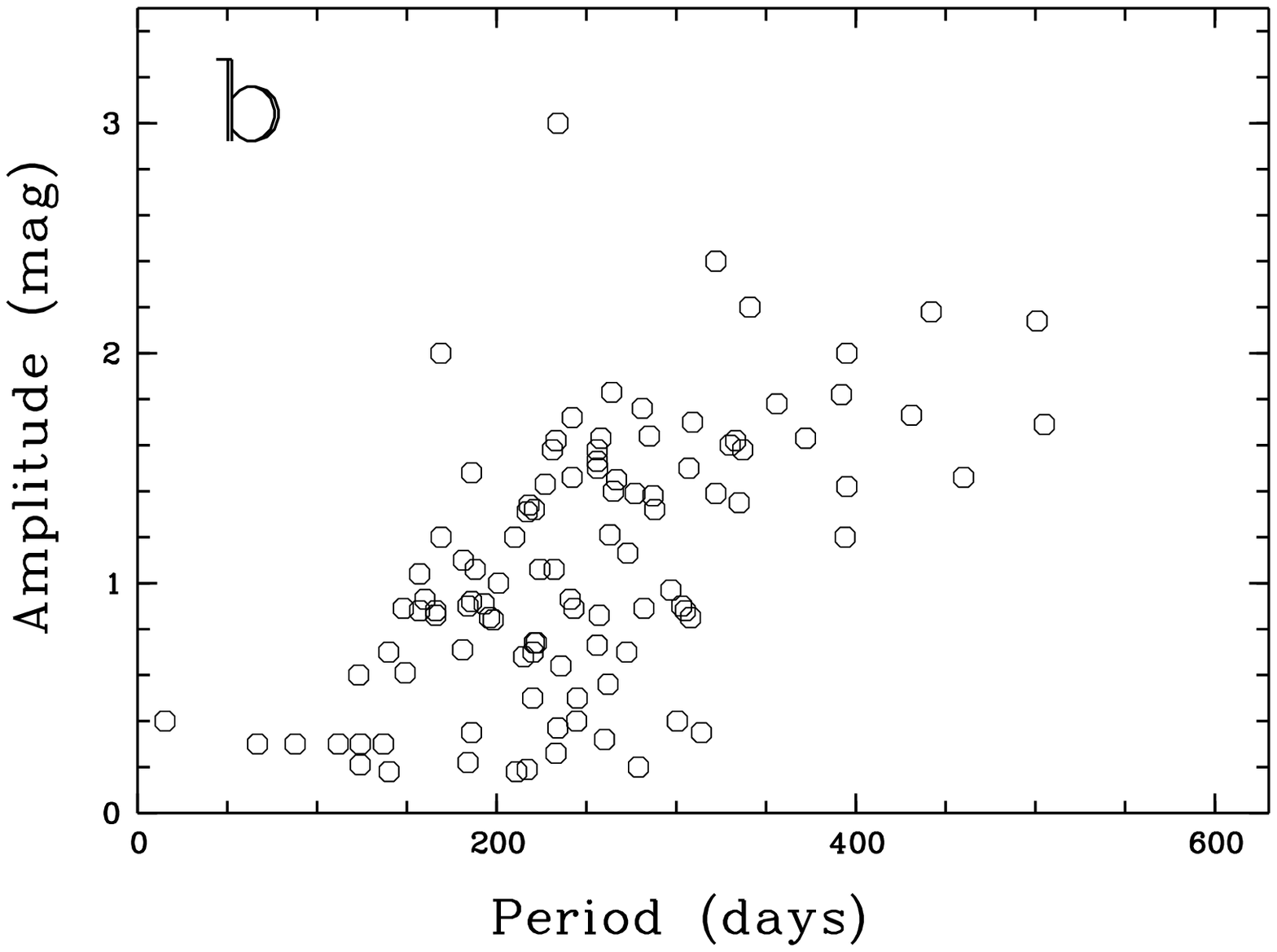}
\caption{Catalina amplitudes  plotted as a function of pulsation periods for MSRa in the halo. All stars plotted here have $|Z|$\,$>$\,$5$\,kpc. We show in panel {\bf a}  our sample stars obeying  $(J-K_{\rm s})_0$\,$<$\,$1.3$. In panel {\bf b}, we show carbon stars. Note the large difference of these two distributions, strongly suggesting that most of the stars in the  upper panel are not carbon rich. All stars plotted here have $|Z| > 5$\,kpc.}
\end{center}
\end{figure}

\subsection{Apparent positions}

 In Fig.~4 (middle panel), we plot M stars verifying $|b| >$ 20$^{\circ}$, and $|Z| >$\,5\,kpc. With these selections, the disc members nearly vanish. Clearly, a very large number of these stars  are located close to the  sinusoid fitted on C stars. There are many more objects with R.A. between 0$^\circ$ and 80$^\circ$, but the most numerous population is in the Sgr leading arm with R.A. between 160$^\circ$ and 250$^\circ$.

\begin{figure*}[!ht]
\begin{center}
\includegraphics*[width=5cm, angle=-90, origin=c]{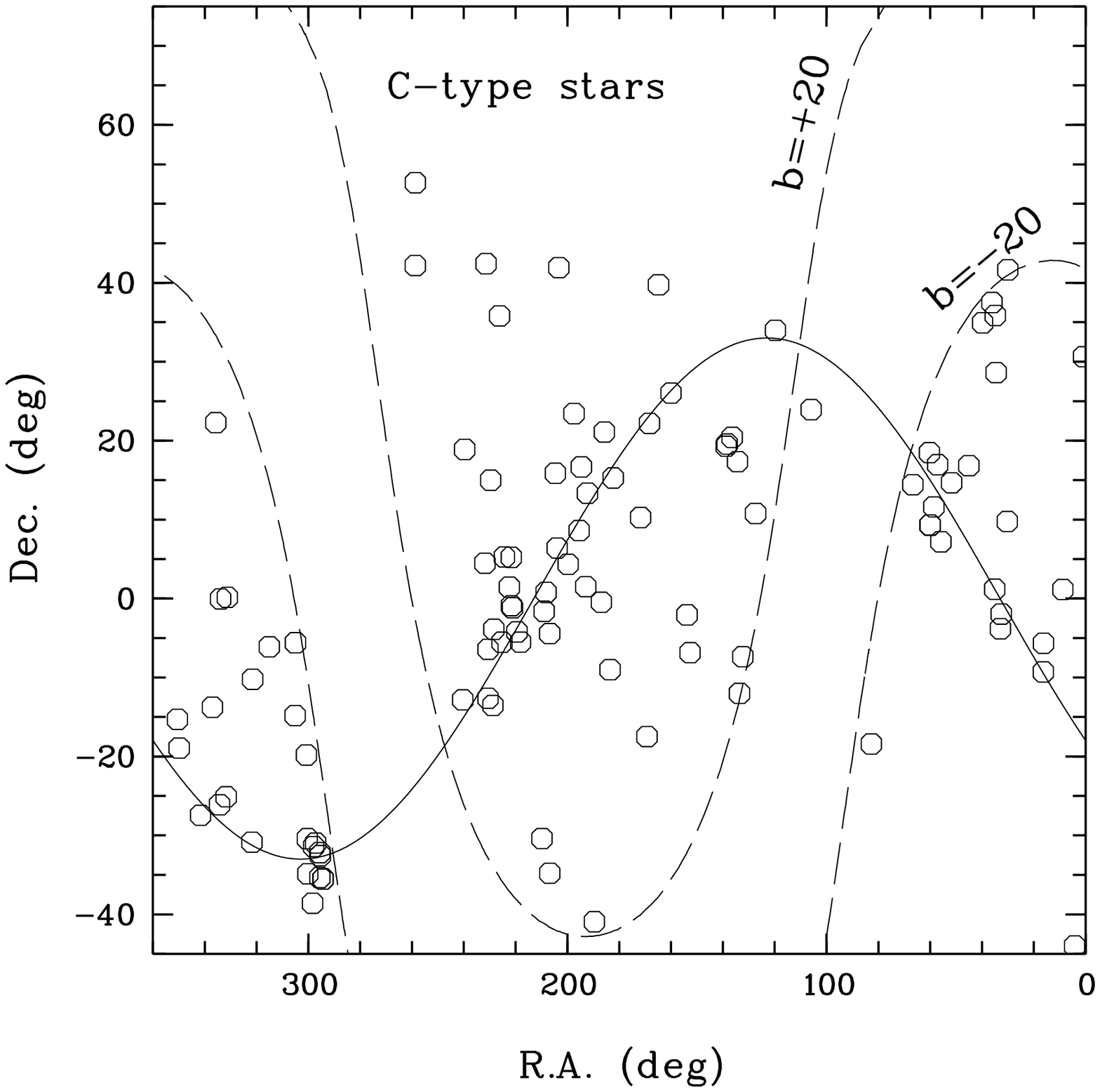}
\includegraphics*[width=5cm, angle=-90, origin=c]{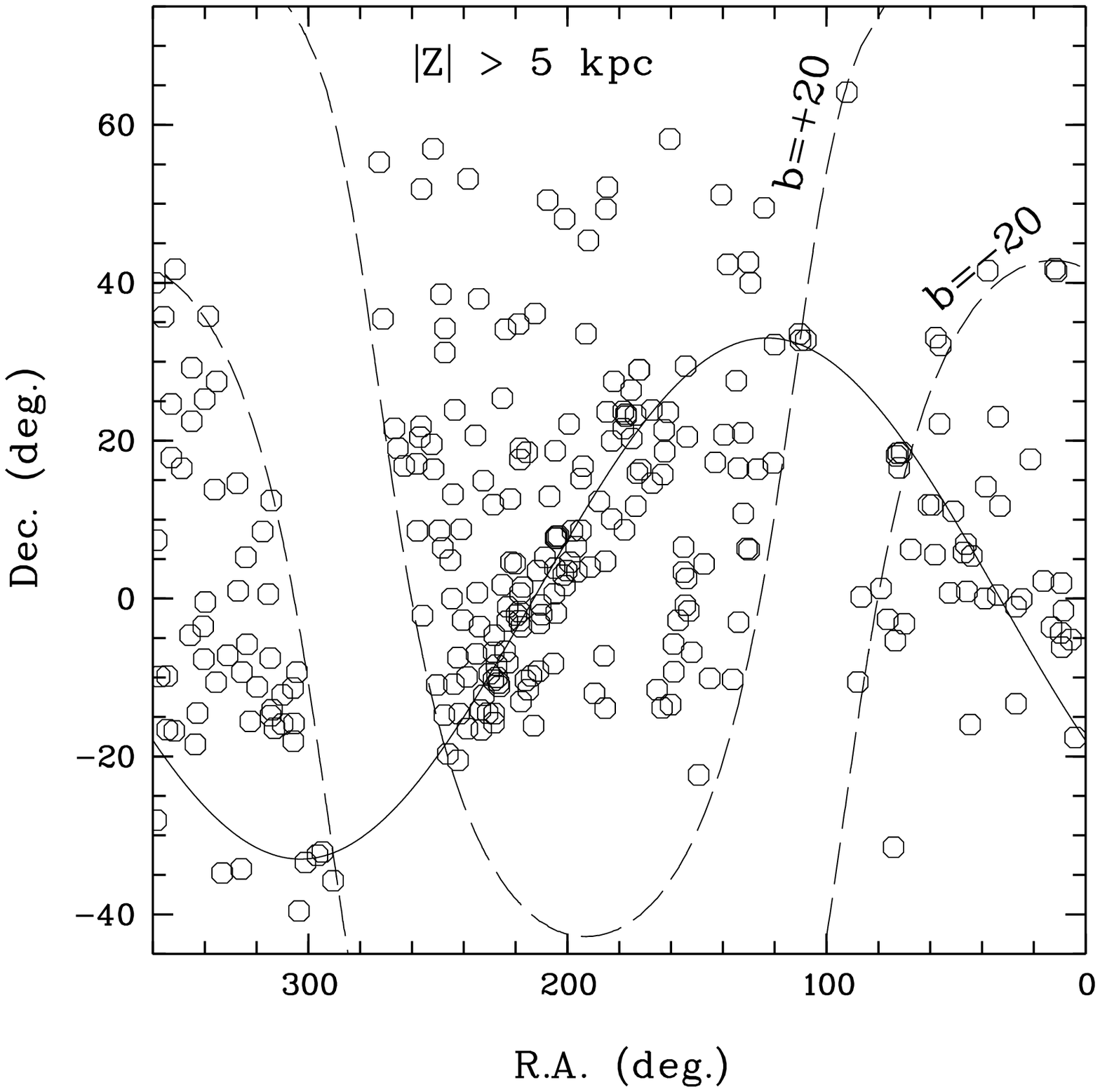}
\includegraphics*[width=5cm, angle=-90, origin=c]{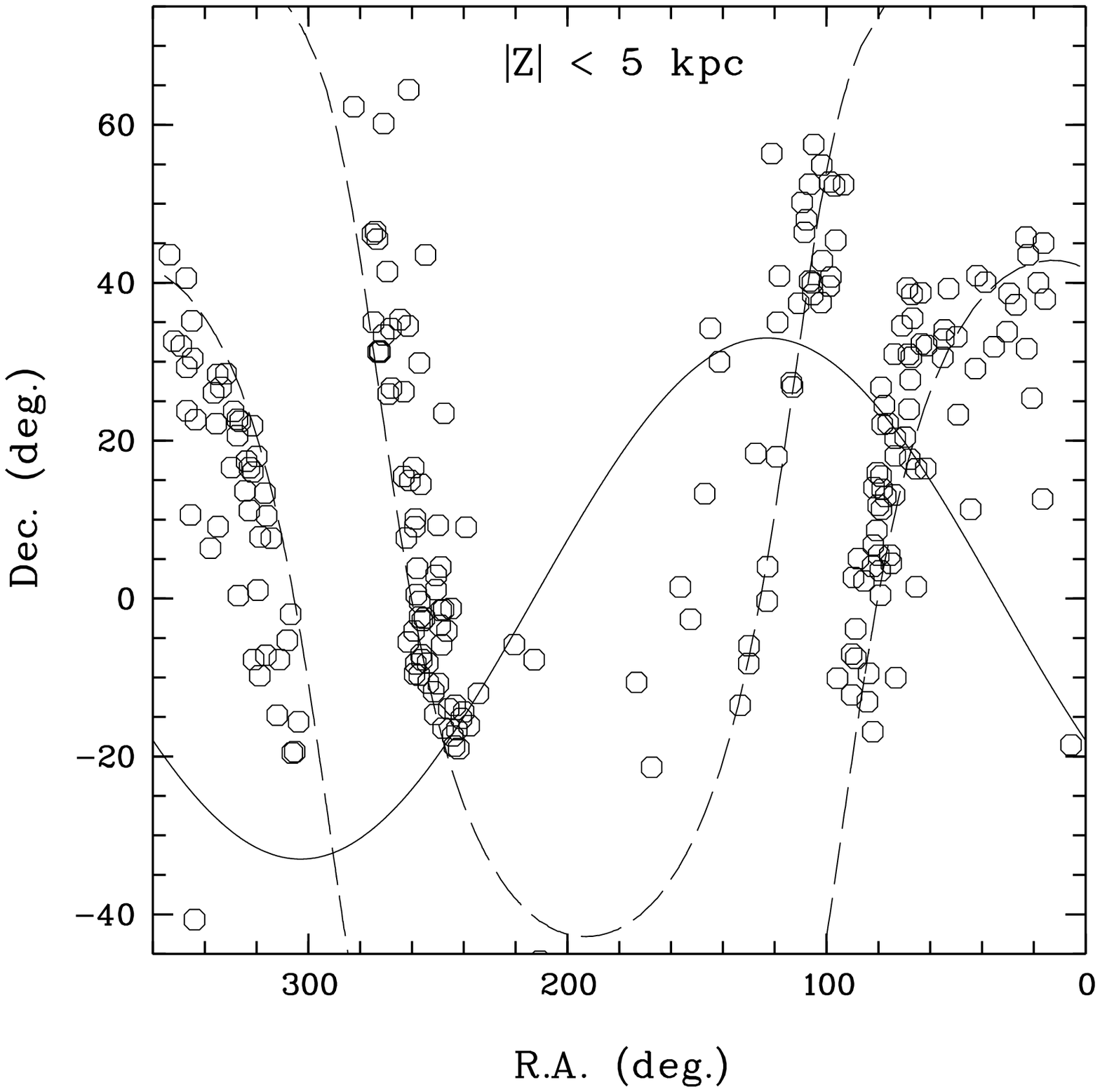}
\caption[] {Apparent positions of pulsating AGB  stars in equatorial coordinates. In the left panel, we plot all known  C stars that are periodic and have $|Z| >$\,5\,kpc.  The sinusoid is an eye fit to the majority of these C stars. In the middle panel, we plot stars of our sample (M type) having $|Z| >$\,5\,kpc. In the right panel, we plot stars of our sample having  $|Z| <$\,5\,kpc. Dashed lines indicate  $|b|=20^{\circ}$. Note that M stars are more numerous than C stars along the Sgr arms.}
\end{center}
\end{figure*}

\subsection{Separating C and C' pulsating sequences}

The MACHO and OGLE surveys (Alcock et al.\ 1997; Udalski et al.\ 1994) monitoring  stars in the Large Magellanic or the Galactic Bulge have shown  that MSRa stars  pulsate in different modes. Several period-luminosity sequences exist, in particular  the so-called sequence C for fundamental pulsation and sequence C' for  first overtone (Wood et al.\ 1999; Ita et al.\ 2004).  Here, in order to avoid confusion with C (carbon) stars, we name seq-C, seq-C', and seq-D the three P-L sequences considered in this work.  In the Magellanic Clouds or the Bulge, the sequence  separation was made possible because all stars were at the same  distance. In the case of stars in the halo or in the Sgr arms, this distance  is  unknown and has to be determined. For carbon stars, HG2015 based their classification seq-C or seq-C' on three diagrams involving the $(J-K_{\rm s})_0$ colour index, periods and amplitudes.  Although, as they note, the diagram log($P$) versus $(J-K_{\rm s})_0$ presents some difficulties to  determine this separation, they reach the conclusion that very few (5 of 121) are seq-C'  stars. 

Concerning our sample stars, none of which are known C stars, we found that the clearest results are obtained by considering the $(I-\KS)_{0}$ colour. We found this colour to be more efficient than $(V-\KS)_{0}$ to separate sequences equally well for  the LMC and the halo sample. 
 Figure~5 shows the situation of LMC stars,  for which sequences are known. Here,   $I$ is a time-averaged magnitude from OGLE and $\KS$ is from the single epoch 2MASS survey. We draw a boundary between seq-C and seq-C' populations. Figures 6 and 7 show the same diagrams for the  stars of our sample, with $|Z| >$\,5\,kpc and $|Z| <$\,5\,kpc, respectively. The $I$-band photometry that we use is  derived from  the Catalina $V_{\rm CSS}$ magnitude and the 2MASS $J-\KS$ colour, as explained  in the Appendix. 

 In  Fig.\,6 and 7, the plotted boundary lines  are identical to those of the LMC. One could object that boundaries separating sequences might change with population characteristics, such as metallicity. Here, we  assume that there is no change of these boundaries. The metallicity of many  giant stars in the Sgr arms is close to  that of the LMC, where [Fe/H] $\approx -0.5$ on average. More precisely, observations of the metallicity distribution in the Sgr arms can be found in  Monaco et al.\ (2007),  Chou et al.\ (2007), and Carlin et al.\ (2018), among others. The distribution peaks at $\sim -0.75$ and has a lower metallicity component. Therefore, it seemed to us reasonable to keep  boundaries unchanged, although we recognize that other factors, like star formation history, could play a role. Certainly, our assumption deserves additional investigation. 

 In Fig.~6, most of our stars are  to the right of the boundary. Six objects are clearly to the left, and we assign them to sequence C'. The position of about 20 objects, to the right of a dotted line, suggests that they are not members of sequence C. Comparing Fig.\,6 and 7 shows that the main cloud shifts to the red and longer periods in Fig.\,7. Since disc stars dominate the $|Z| <$\,5\,kpc population, one interpretation of this shift is that it arise from the higher metallicity of the disc as compared to the   halo. In Fig.~7, we consider that only five stars are separated enough from the boundary line to be considered as being on the sequence C'. Three stars may lie on the D sequence.

The fact that so few stars are found in sequence C' is puzzling, in strong contrast with what is seen in the LMC. We first note that exactly the same fact occurs for halo carbon stars, as found by HG2015. Additionally, the HG2015 method of separating sequences and our method are different, leading however to similar conclusions. HG2015 suggested that seq-C' pulsators have    amplitudes too small to be included in the Catalina MSRa catalogue of Drake et al.\ (2014).  This may be true, although not entirely convincing because $\sim$ 15\% of a total of 1200 LMC seq-C' stars have a $V$-band amplitude larger than 0.55~mag  in the OGLE catalogue. Such a large amplitude is relatively well detected by Catalina. This point remains unexplained for us.

\begin{figure}[!ht]
\begin{center}
\includegraphics*[width=7cm, angle=-90]{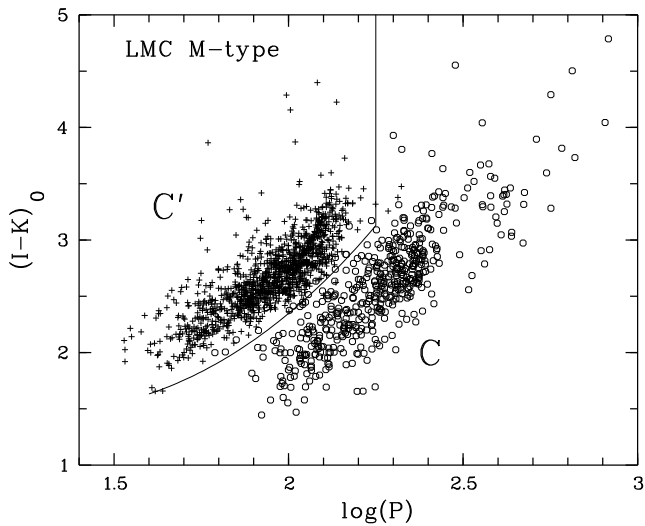}

\caption[] {Plot of the $(I-\KS)_0$ colour index versus period (in days)  for oxygen-rich  MSRa variables of the LMC. 
The curved line  separating sequence C and C'  is $y = a x^2 + b$ where $x = log(P) - 1.2$, $a = 2.3$, $b=1.45$. The vertical line corresponds to $P = 178$ days.}
\end{center}
\end{figure}

\begin{figure}[]
\begin{center}
\includegraphics*[width=7cm,angle=-90]{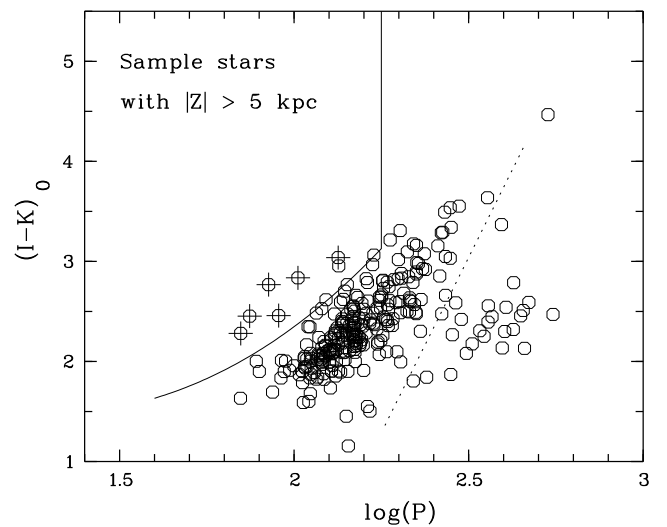}
\caption[]{Plot of $(I-\KS)_0$  $versus$ period (in days) for our sample  stars located at more than 5~kpc from the galactic plane.
The curved line is the same as for LMC. Stars with an overplotted cross are attributed to sequence C'. 
Stars to the right of dotted line are considered to belong to sequence D.}
\end{center}
\end{figure}

\begin{figure}[]
\begin{center}
\includegraphics*[width=7cm,angle=-90]{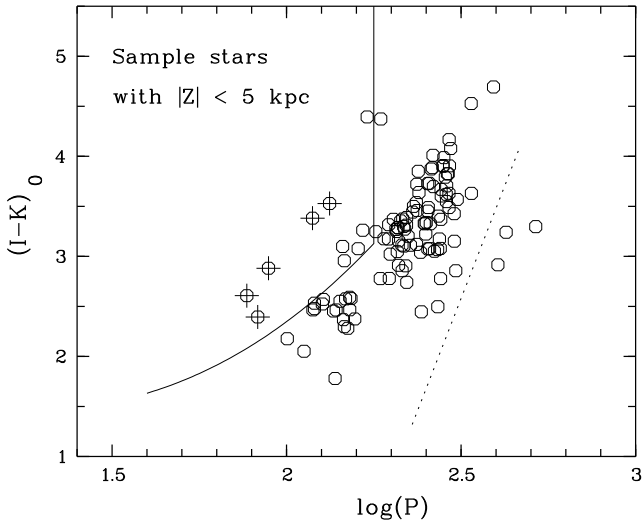}
\caption{Same as Fig.~6 for stars of our sample  being closer than 5 kpc from the galactic plane. }
\end{center}	
\end{figure}

%==============================================================================================

\section{Results}

\subsection{Distances}

After having determined the pulsation sequences for stars of our sample, we can derive their distances. For that purpose, we use, as HG2015 did for C stars, the near infrared Wesenheit magnitudes $W_{JK}$. We use the P-$W_{JK}$ relations given in Table 1 of Soszy\'{n}ski et al.\ (2007) for O-rich sequences labeled C$_{\rm O}$, C'$_{\rm O}$ and D$_{\rm O}$. We adopt $m-M = 18.49$ for the LMC distance (e.g. de Grijs et al.\ 2014). 

The uncertainty on distances is made of:\, a) the measurement error of the $K_{\rm s}$ 2MASS magnitude provided in the 2MASS catalogue, of order 0.03\,mag; b) the relative uncertainty on the period found to be $\sim$~2\%, from comparing  periods given by LINEAR and those of Catalina, implying 0.04 mag on the absolute magnitude;  c) the scatter of the P-L relation, i.e. $\sigma = 0.15$~mag; and finally d) the uncertainty that originates in the single epoch character of \KS. Ideally we should use  a time-averaged \KS, but this quantity is lacking.  This implies an uncertainty amounting the \KS-band amplitude that we estimate as explained below. 

  For C stars, HG2015 exploit a   correlation between the $K_{\rm s}$ amplitude and the $(J-K_{\rm s})_{\rm0}$ colour, established for dusty Miras with that colour lying between $\sim$1.5 and 5. They use an extrapolation to the bluer colour for many C stars. Our stars are M type, and almost none are as red as that. We  adopt another approach based scaling the observed $V$-band 
amplitude $\Delta V.$ We compare this quantity with the difference between $K_{\rm s}$ from 2MASS and $\KS$ from the DENIS  survey (Epchtein et al.\ 1999). These surveys provide  independent data taken at two different dates.  DENIS and 2MASS have very similar $K$-band photometries: according to  Carpenter (2001), $\KS$(DENIS) is  smaller than $\KS$(2MASS), by $\sim 0.02$~mag for the range of $(J-K_{\rm s})_{\rm0}$ of our stars.  Therefore, we expect that, for a given $\Delta V$, the $\KS$ difference can reach  the  $K_{\rm s}$ amplitude (all along this work,  amplitudes  mean peak-to-peak amplitudes). 

By cross-matching our sample with the DENIS catalogue and requesting $K_{\rm s} >5$ to avoid saturation results in  136 stars. Figure~8 shows this approach to work reasonably well.  The average shift between the two photometries was found to be 0.05 mag. In Fig.\,8, the error bars drawn in ordinates are derived from summing in quadrature the 2MASS and DENIS catalogue errors. On abcissa, the following errors that we believe reasonable were adopted: $\pm$0.15\,mag when $\Delta V =$~2.5 mag down to $\pm$0.05\,mag if $\Delta V =$~0.3. This plot shows  that $ \Delta \KS \approx 0.40 \times \Delta V$.  For example, when $\Delta V =$~2 mag, Fig.\,8 shows that the absolute value between  the two $\KS$ values can reach 0.8 mag.

After including the $\sim$~0.2 mag error discussed in Sect.\,3.1, the last uncertainty term (called d above) can reach $ 0.5 \times (0.4 (\Delta V + 0.2) )$ mag. Typically, $\Delta \KS\sim$ 0.1 mag for $\Delta V = 0.3$, and $\Delta$\KS\, $\sim$\,0.4~mag. for $\Delta V$~$=$\,2.  The two last uncertainties (named c and d above) dominate, and distances are derived to within  20\% or better.

\begin{figure}[]
\begin{center}
\includegraphics*[width=5cm,angle=-90]{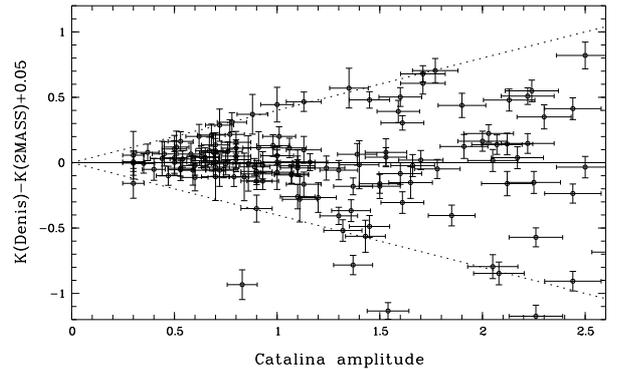}
\caption{Difference between $K_{\rm DENIS}$ and  $K_{\rm 2MASS}$ as a function of the Catalina amplitude for 136  stars of our sample obeying $K_{\rm 2MASS} > 5$.  This diagram  suggests the $K$-band amplitude  to be around 0.4 $\times$ the $V$-band Catalina amplitude, as indicated by dotted lines.}
\end{center}	
\end{figure}

The derived distance can be large, up to 150 kpc. In Fig.~9, we plot the distances  as a function of the longitude \LAM  in the plane of the Sgr orbit. Coordinates in this plane \LAM and \BET are derived from Appendix A of Belokurov et al.\ (2014).  Here, we consider objects with $ |\BET| < 13$\degree. The continuous line traces the mean locus of RR Lyr stars as given by Hernitscheck et al.\ (2017). The 1$\sigma$ depth of the RR Lyrae location is about 4\,kpc and 6\,kpc for the leading and trailing arm, respectively. For the leading arm, there is reasonable agreement between RR Lyrae and M stars. For example, near \LAM $=$~ 50\degree ~and 90\degree, agreement is very good. However, for $\LAM = 60$-$70$\degree, many of our M stars are slightly farther away from us. Concerning the trailing arm, the distances are in agreement for 70\% of the stars, but are larger (by up to 40\%)  for 30\% of them.

 At $\LAM \sim 120^{\circ}$, we indicate two stars, MMK-153 and MMK-154, at distances 92$\pm$18~kpc and 98$\pm$13 kpc, respectively.  Coordinates and some photometric information of stars discussed in the text are provided in Table 2. MMK-153 and MMK-154   belong to the globular cluster Pal~4 and were already known. According to Harris (1996, 2010 edition), this cluster is at 108~kpc, with a metallicity Fe/H$ = -1.41$. At distances of 100 kpc, they represent excellent examples on which other stars can be compared. In particular, obtaining for these two M stars the Pal~4 distance within uncertainties  supports our method for distance determinations.

\begin{table}[!h]
\caption[]{ Coordinates (J2000) from 2MASS, deredenned 2MASS \KS magnitudes, and periods (from Catalina) for MMK stars cited in the text}
\begin{center}
\begin{tabular}{lrrrr}
\noalign{\smallskip}
\hline
\hline
\noalign{\smallskip}
 Star & R.A (deg) & Dec (deg) & $K_0$ & $P$(days)  \\
\noalign{\smallskip}
\hline
\noalign{\smallskip}
MMK-003 &   8.55187 & -1.53182 & 14.15 & 142\\ 
MMK-007 &  13.15590 & -3.61278 & 13.85 & 160\\
MMK-092 & 108.00498 & 32.71524 & 13.39 & 127\\
MMK-094 & 110.00952 & 32.70439 & 13.32 & 109\\
MMK-095 & 110.47729 & 33.50544 & 12.66 & 224\\
MMK-101 & 120.07031 & 32.16365 & 13.80 & 134\\
MMK-114 & 132.35449 & 20.98200 & 13.82 & 297\\
MMK-153 & 172.30163 & 28.97090 & 13.06 & 150\\
MMK-154 & 172.31129 & 28.97068 & 13.59 & 130\\
\noalign{\smallskip}
\hline
\noalign{\smallskip}
\end{tabular}
\end{center}
\end{table}

In the upper right corner of Fig.\,9, we have labelled MMK-003 and MMK-007, because they are among the most distant stars in our sample. They are angularly separated by only 5.7\degree. Interestingly, they are also close to two carbon stars, renamed HG3 and HG5 by HG2015. The former is at 2.6\degree\, from MMK-003, and the latter is at 6.5\degree\, from MMK-007. HG2015 attributed HG3 to the  A16 halo substructure, while this was less sure for HG5. Given  uncertainties,  the distances of our M stars, 145 $\pm$ 20 kpc,  are in remarkable agreement with  those of the carbon stars, 114 $\pm$ 15 and 140 $\pm$ 22 kpc. The light curves of the eight more distant stars shown in Fig.~9 are given in Fig.~10. Periods are those given by the Catalina DR2 database. The two light curves plotted at bottom of Fig.~10 are MMK-153 and MMK-154, members of Pal~4.

\begin{figure}[]
\begin{center}
\includegraphics*[width=8cm,angle=-90]{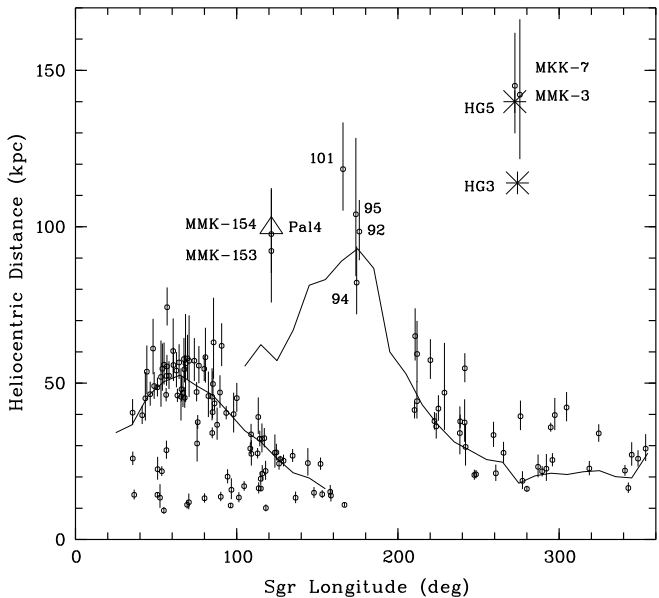}
\caption{Heliocentric distance as a function of the Sgr longitude. Only stars  within 13 degrees from the Sgr orbit plane are plotted. The continuous line is the average position of RR Lyr variables of the Sgr arms given by Hernitschek et al.\ (2017). The 1$\sigma$ scatter of this RR Lyrae positions is $\sim$~5~kpc. The globular cluster Pal~4 is indicated by a triangle. Two carbon stars, HG5 and HG3, are plotted with asterisks.}
\end{center}	
\end{figure}

\begin{figure}[]
\begin{center}
\includegraphics*[width=3.cm,angle=-90]{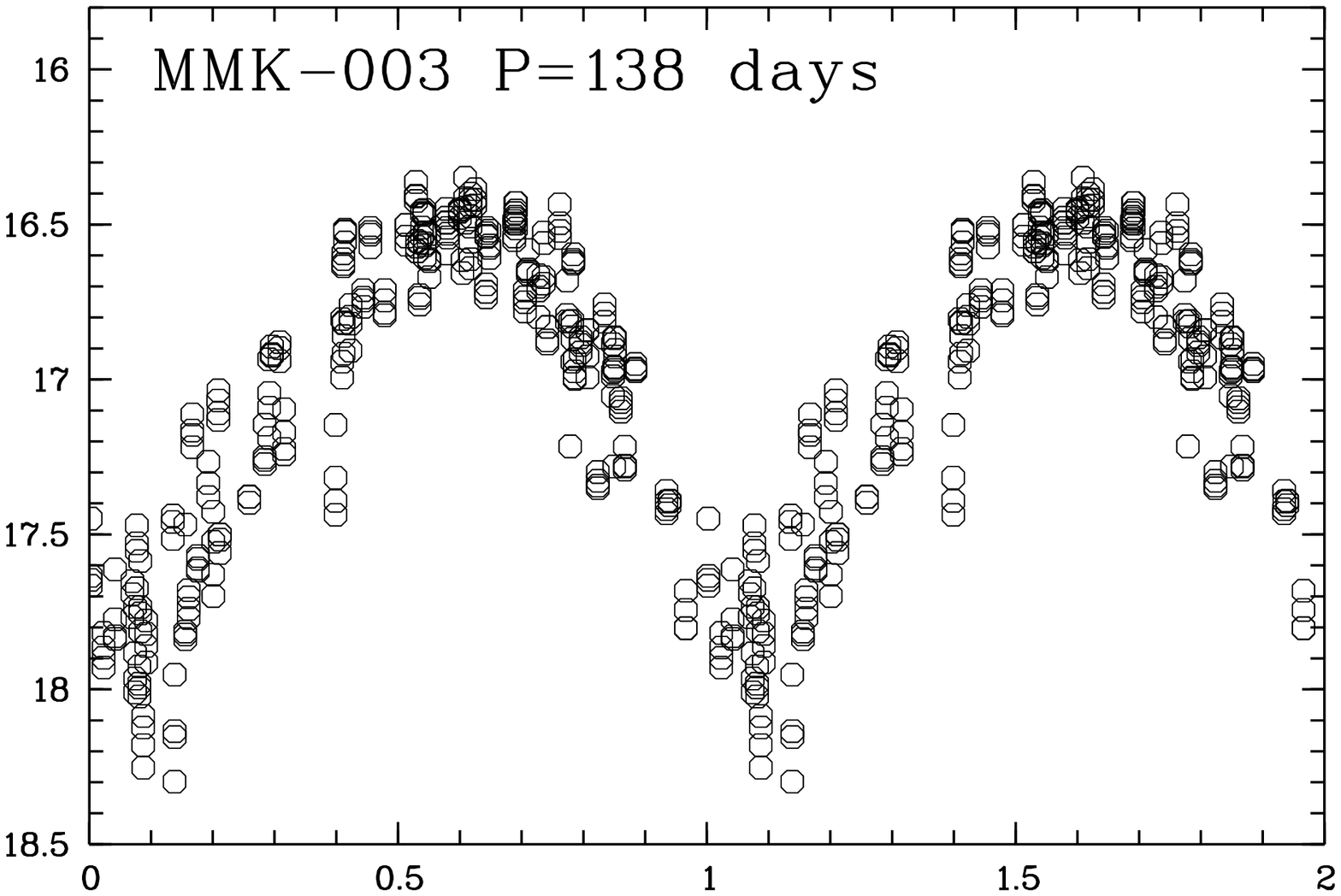}
\includegraphics*[width=3.cm,angle=-90]{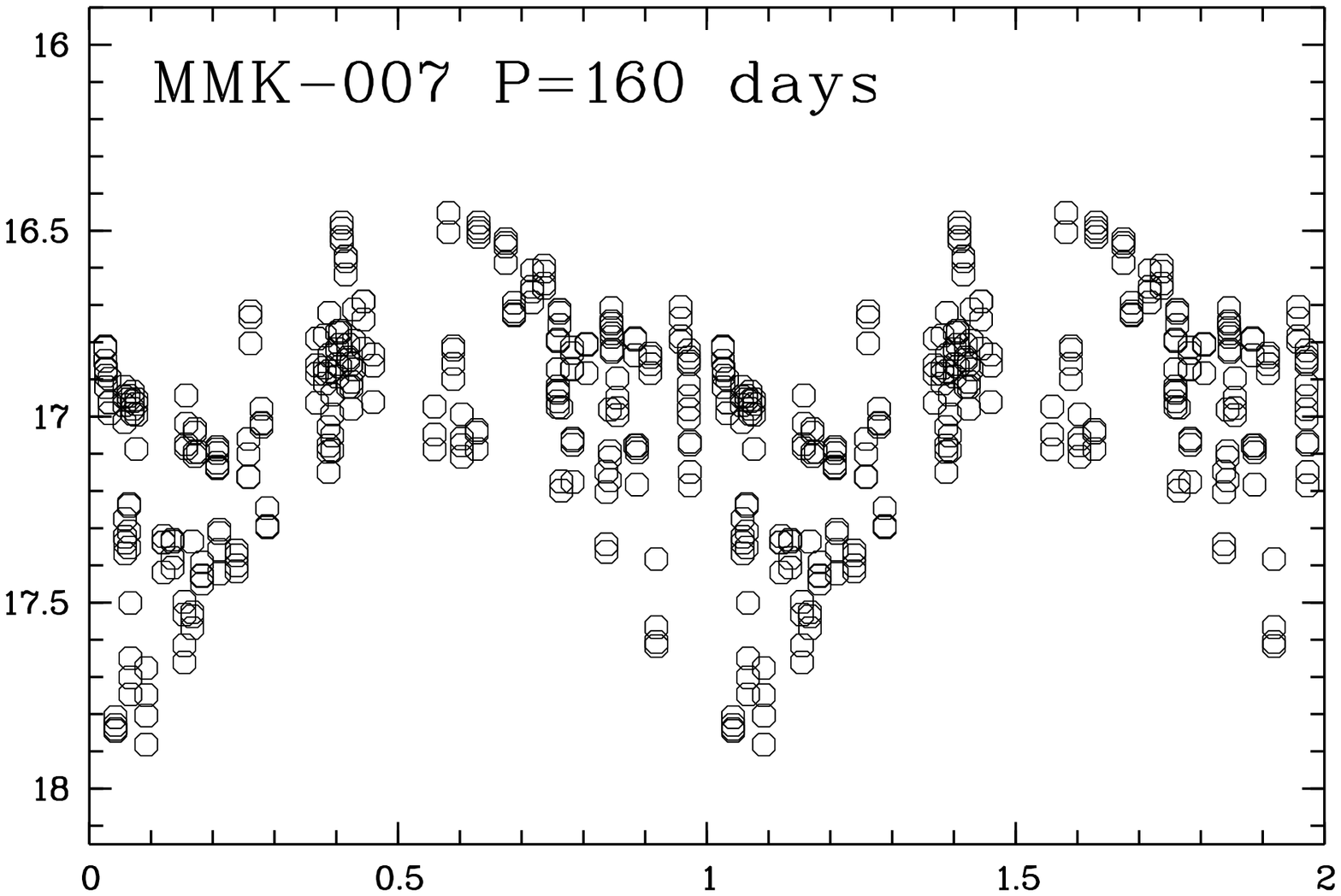}

\includegraphics*[width=3.cm,angle=-90]{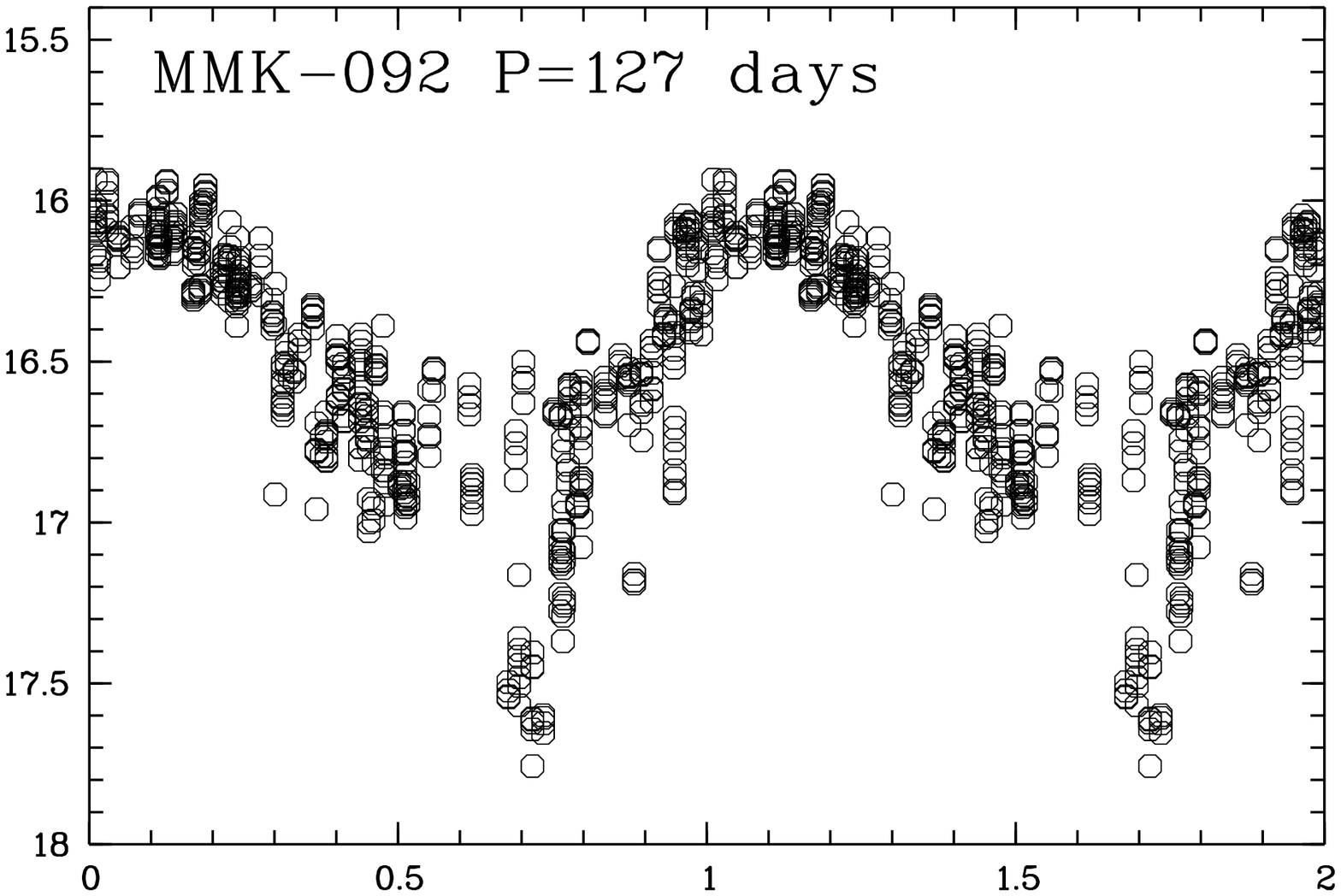}
\includegraphics*[width=3.cm,angle=-90]{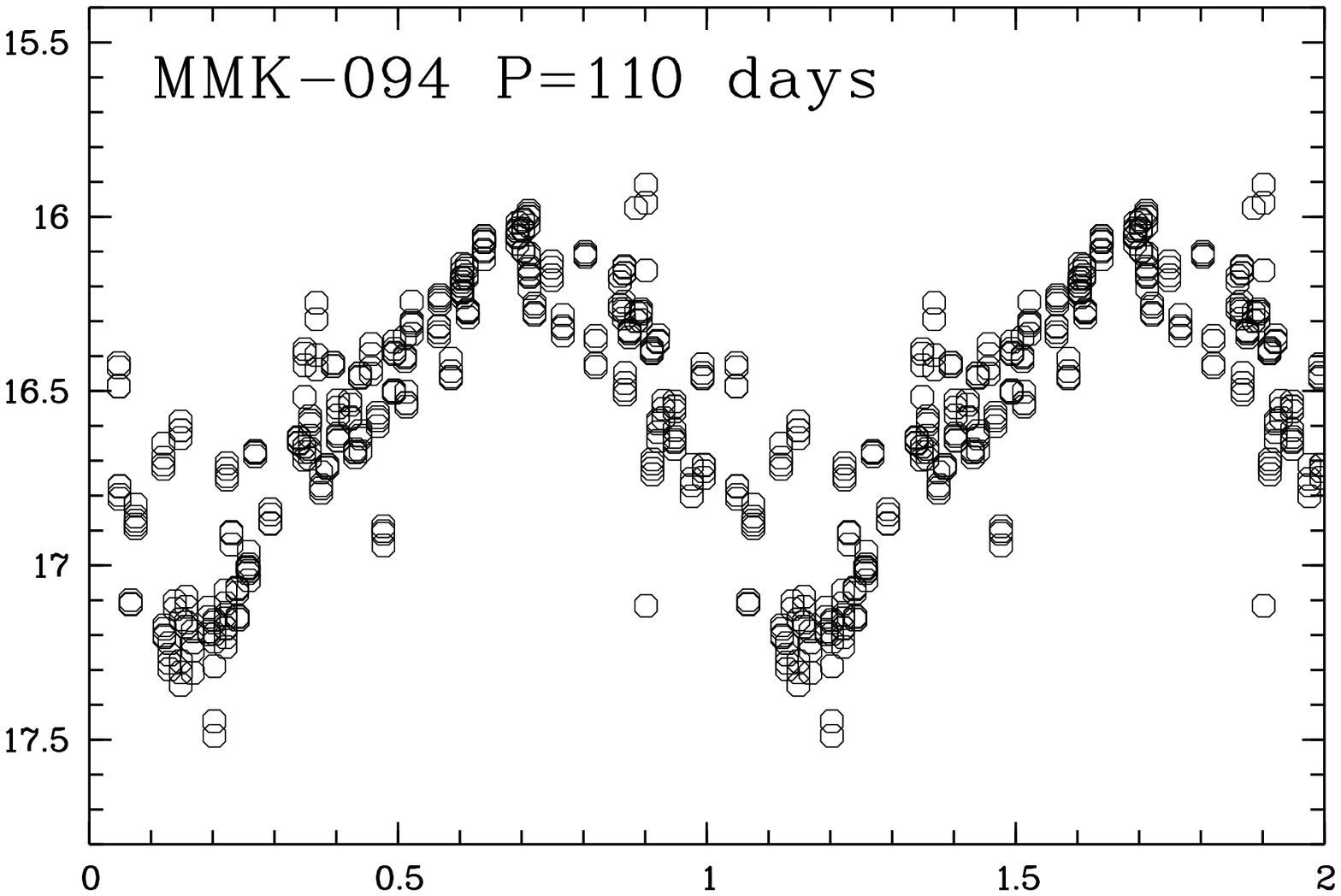}

\includegraphics*[width=3.cm,angle=-90]{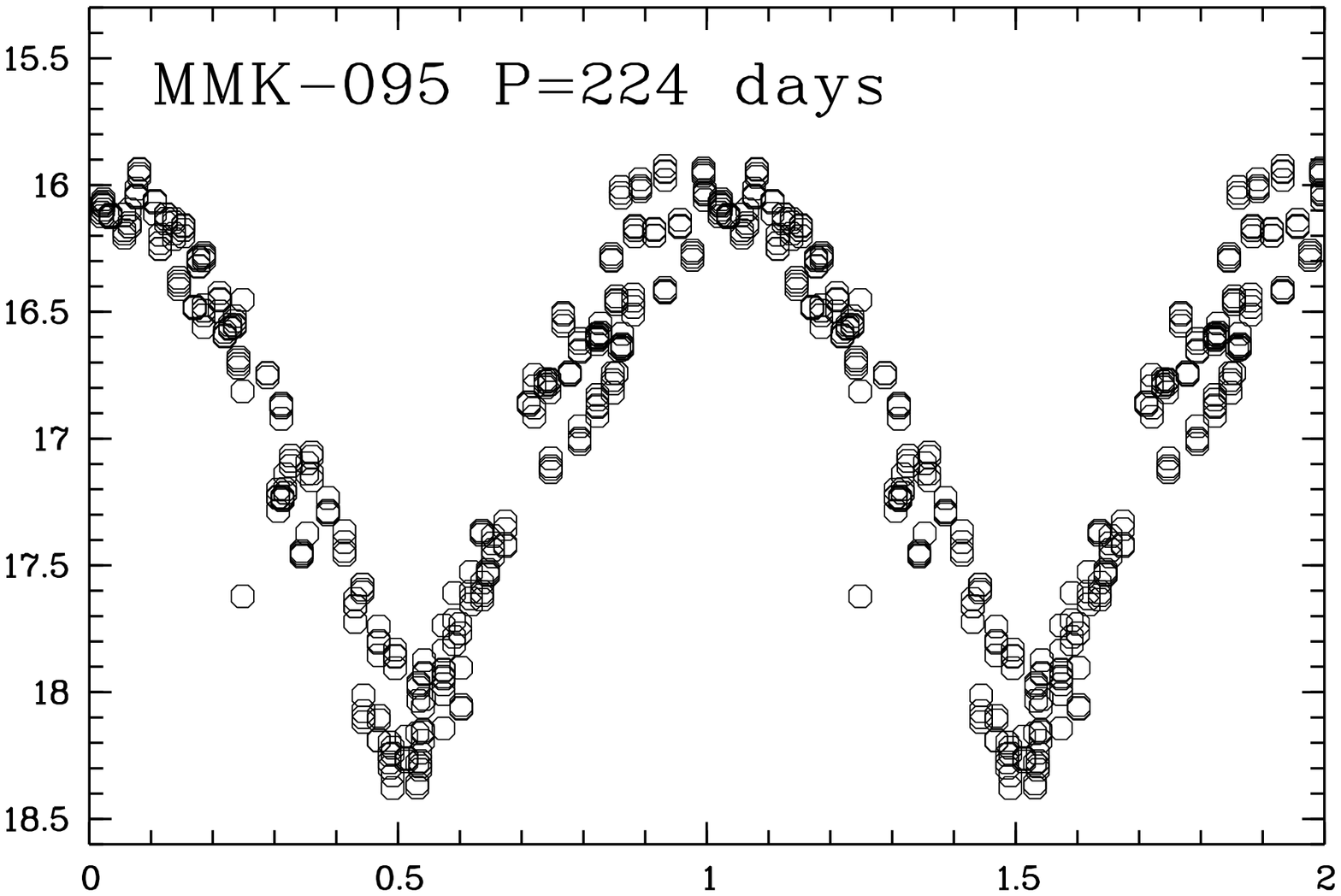}
\includegraphics*[width=3.cm,angle=-90]{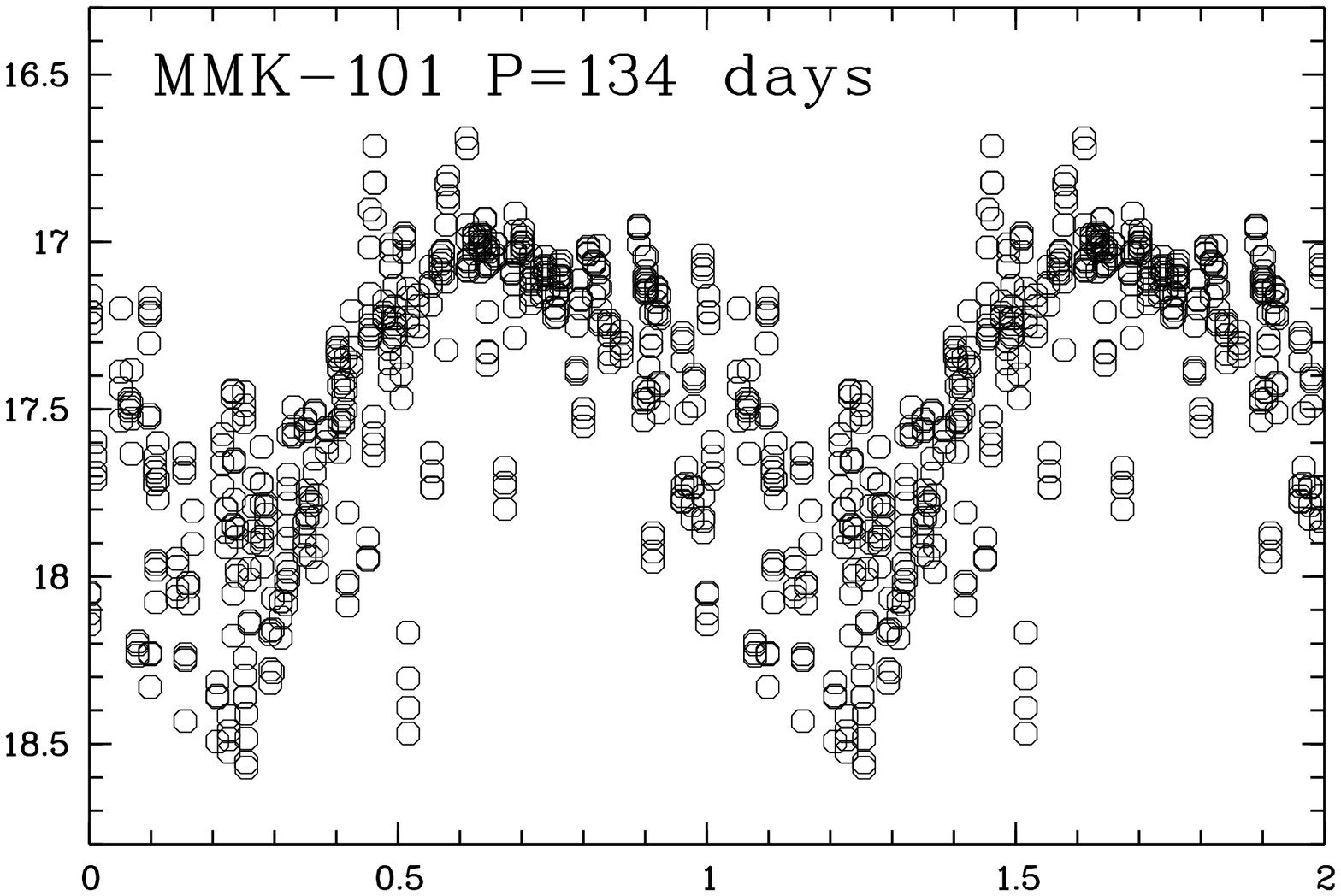}

\includegraphics*[width=3.cm,angle=-90]{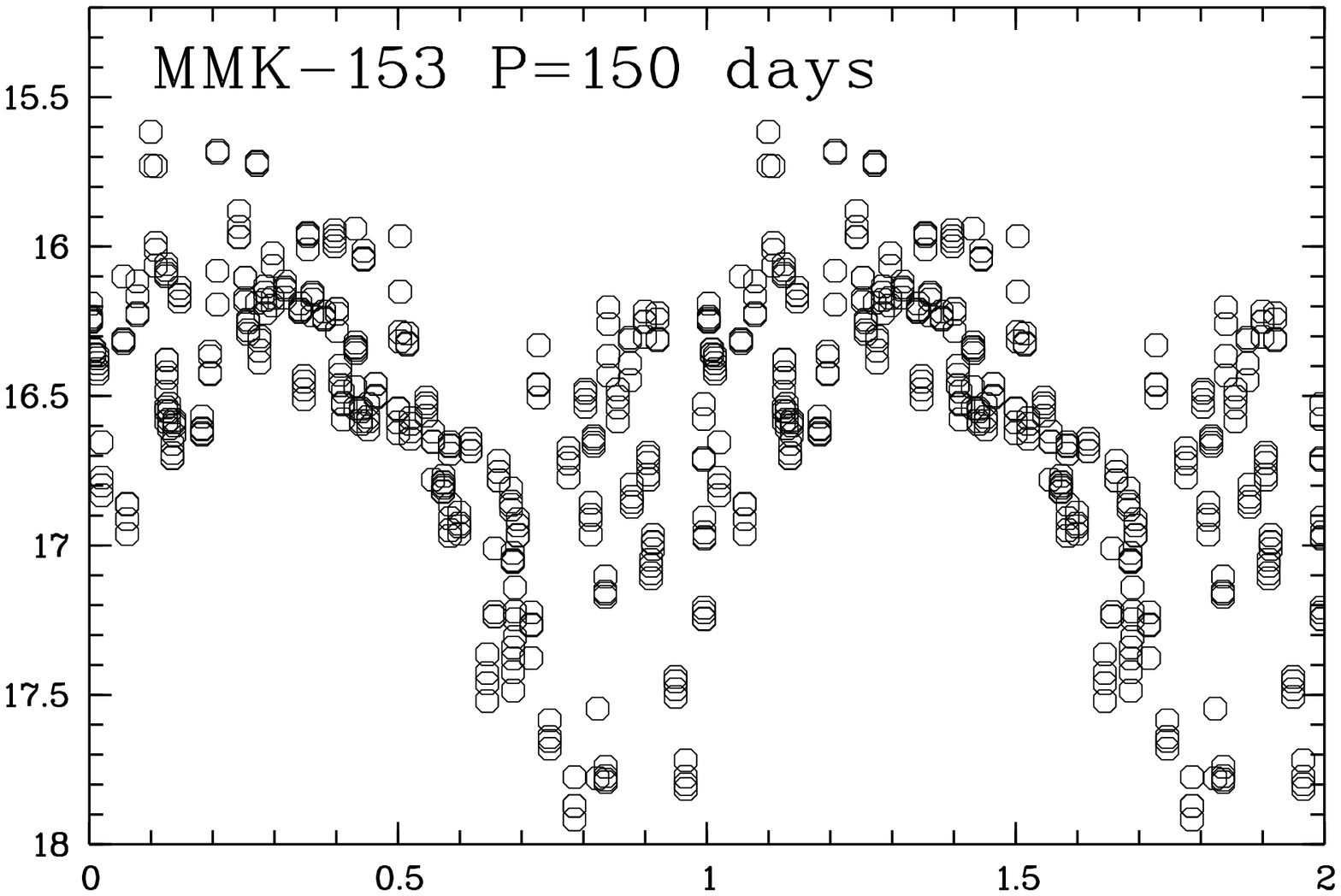}
\includegraphics*[width=3.cm,angle=-90]{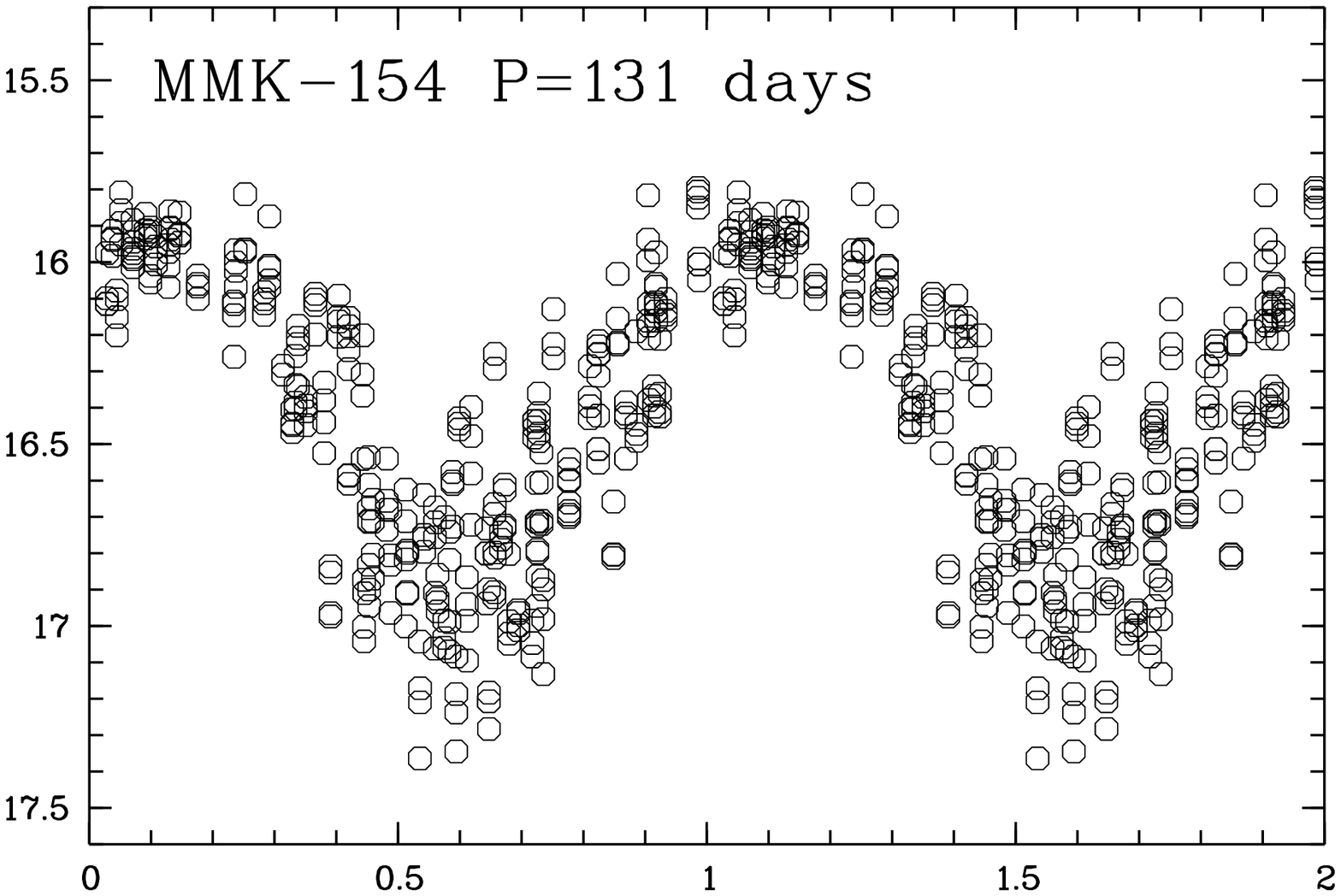}

\caption{Plots of Catalina magnitude as a function of phase, for labelled stars in Fig.~9. The lightcurves at bottom refer to two members of the Pal~4 globular cluster.}
\end{center}
\end{figure}

\subsection{A peculiar clump}

When one looks at Fig.~4, middle panel, one can see that, south of this line, there exists a group of stars  at R.A.\,$\sim$\,160$^{\circ}$. This part of the sky is zoomed in Fig.~10, where we plot M stars (circles), known C stars (filled circles), together with the two stars of pair $6$ of Starkenburg et al.\ (2009). Crosses indicates those objects at more than 25 kpc.  The three C stars are, from bottom to top, m7 (alias HG52), m36 (alias HG 48), and m04 (Mauron et al.\ 2004). The last one has been missed in the census of HG2015. The distances and heliocentric radial velocities $V_{\rm r}$ of these three C stars are 40\,kpc \& $+342$\,\kms , 24\,kpc \& $+$306\,\kms, and 75\,kpc \& $+$202\,\kms, respectively. As already noted by HG2015, the first of these three stars may have a link with the Starkenburg et al. stars located at 30\,kpc with $V_{\rm r} = $\,380\,\kms. The other two C stars are isolated. We have also indicated the position of the Orphan stream which is well detected by a variety of methods (Grillmair 2006; Belokurov et al.\ 2006; Bernard et al.\ 2016) and is at a distance of $\sim$ 20\,kpc. It is clear that our clump is not the Orphan stream. Radial velocities  are necessary to  confirm or deny the physical reality of this peculiar feature. 

\begin{figure*}[]
\begin{center}
\includegraphics*[width=6cm,angle=-90]{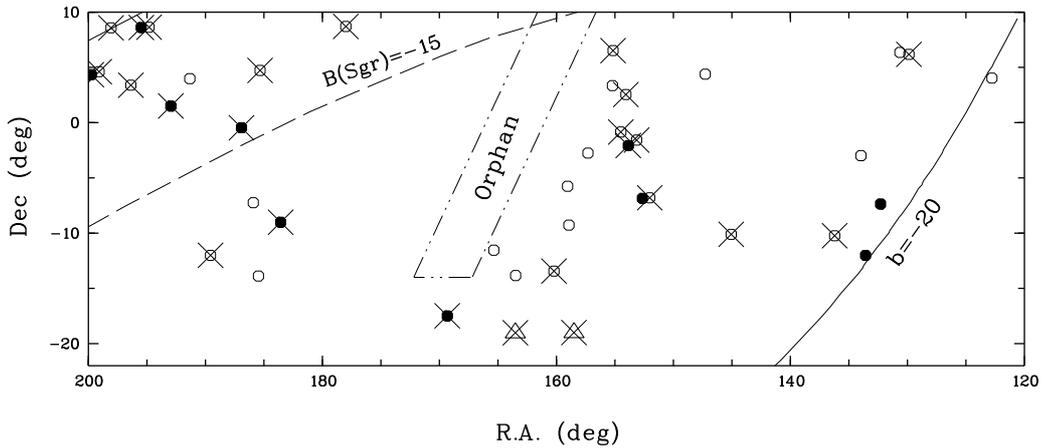}
\caption{Close-up on the region of the group of stars seen at R.A. $\sim$ 160\degree. M stars and C stars are plotted with  open and filled circles, respectively. Two triangles at bottom are  pair 6 of Stackenburg\,(2009). Overplotted crosses are stars more distant than 25~kpc from the Sun. Those at the upper left corner belong to Sgr arm. The Orphan stream position is indicated.}
\end{center}	
\end{figure*}

\subsection{{\it Gaia} data}

\subsubsection{Parallaxes}

In the {\it Gaia} DR2 (Gaia collaboration et al. 2016), parallaxes are  available  for all the sample  stars but two. However, half of them are negative and the positive ones have large uncertainties. In the whole sample,  only 7 stars have $|\epsilon(\pi)/\pi)| < 0.2$.  Therefore, it is necessary to wait for more robust Gaia measurements and future data releases in order to compare them to our distances based on period-luminosity relations.

\subsubsection{Radial velocities}

 We  first collected all {\it Gaia} radial velocities. We added those from the APOGEE (Majewski et al.\ 2017) and RAVE (Steinmetz 
et al.\ 2006; Kordopatis et al.\ 2013) surveys, and a few from Monaco et al.\ (2007). There is in most cases excellent agreement between them when two measurements are available, with differences of $\leq 2$\,\kms. Our catalogue includes these velocities, their errors and references. Some additional velocities come from Feast \& Whitelock (2000), and from Smak \& Preston (1965), with errors $\sim$ 5\,\kms. Although  less accurate, $\sim$ 12-15\kms,  a few velocities come  from the LAMOST database (Luo et al.\ 2016) or the Sloan DR7 M dwarf catalogue (West et al.\ 2011). Overall, of 429 stars on our catalogue,  90 have a radial velocity.

\subsubsection{Proper motions}
$Gaia$ DR2 proper motions allow to discard some distant stars, because these proper motions would imply an excessive tangential velocity ($V_{\rm t}$) (we attribute a negative flag in our catalogue; see below). The relevant relation is  $V_t$\,$=$\,$4.7 \mu \times D$~\kms ~with $\mu$ in mas/yr and $D$ in kpc. To illustrate this point, two of the most distant stars in the Galaxy were found by Bochanski et al.\ (2014), namely ULAS J001535.72$+$015549.6 and ULAS J074417.48$+$25253233.0, at proposed distances of $\sim$\,275~kpc, and $\sim$\,240~kpc, with uncertainties of $\sim$ 70 kpc. These distances are based on the result that stars would be  M-giants, and not dwarfs. Both are interestingly close to the Sgr mean orbit plane: $\BET = +16.3$\degree\, and $\BET = -4.7$\degree, for the first and second  stars, respectively. While the second object has no significant {\it Gaia} proper motion, the first displays $\mu_\alpha cos(\delta)$\,$=$\, $-5.00 \pm 1.11$ mas/yr and $\mu_\delta = -4.91 \pm 0.82$~mas/yr. Were its distance of 240~kpc correct, it would yield a tangential velocity of $\sim$ 9000 \kms. It seems to us more probable that it is a much closer dwarf star, although considerable care was dedicated by Bochanski et al.\ (2014) to spectroscopically discriminating giants and dwarfs. 

Another very distant object was proposed by Mauron et al.\ (2018), named MMK-114 here, with \KS $= 13.8$ and $(J-K_{\rm s})_0 = 1.39$, $B_{\rm Sgr} = -9.2$\degree. The colour makes it unclear to classify  it  M  or C. Its light curve (Fig.~11) is compatible with a semiregular variable having a peak-to-peak amplitude of 0.6 mag. Catalina DR2 provides a tentative period of 297 days. Because it is faint, adopting this period would yield a distance of $\sim$~200~kpc, putting it outside of the frame of Fig.~9. {\it Gaia} DR2 does not provide a significant parallax but gives $\mu_\alpha cos(\delta) = -0.22 \pm 0.26$~mas/yr and $\mu_\delta = -2.1 \pm 0.19$~mas/yr. Provided this $\delta$ motion is confirmed, this star is  closer than 200~kpc in order, again, to avoid excessively large velocity ($\sim$ 2000 \kms). 

The stars MMK-003, MMK-007, MMK-101, MMK-153, MMK-154 have no  significant proper motion. Those close to the apocenter of the trailing arm, MMK-92, MMK-094, MMK-101  have plausible $V_{\rm t} = $ 400, 240, and 270 \kms~at their respective distances. We also checked the eight carbon stars stars listed by Deason et al.\ (2012) (their Table ~2), and found that only J1446-0055 displays some motion at more than  3$\sigma$: $\mu_\alpha cos(\delta) = -1.00 \pm 0.12$~mas/yr and $\mu_\delta = -0.53 \pm 0.10$~mas/yr. At its distance of $\sim $ 80~kpc, this yields a plausible $V_{\rm t} = 425$ \kms.

\begin{figure}[]
\begin{center}
\includegraphics*[width=7cm,angle=-90]{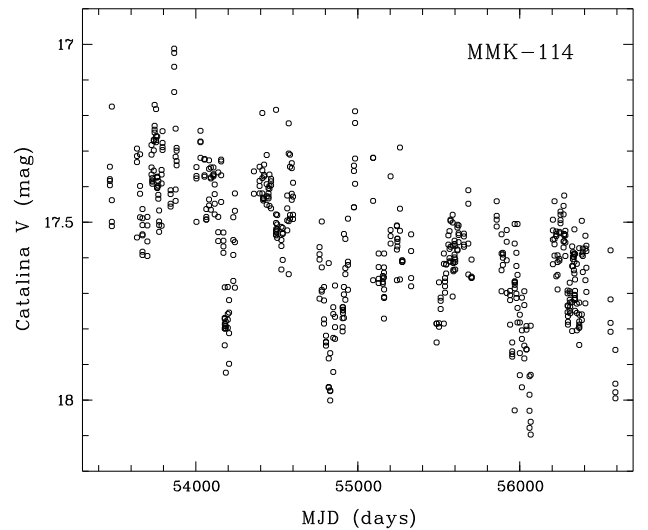}
\caption{Light curve of MMK-114 (see text).}
\end{center}	
\end{figure}

\subsection{The catalogue}

Our catalogue includes the MMK number, $\alpha$, $\delta$ (J2000), $l$ and $b$,   
Sgr orbit coordinates \LAM and \BET,    $E(B-V)$, $(K_{\rm s})_0$, $(J-K_{\rm s})_0$, absolute magnitude $M_{\rm K_s}$, distance and its error in kpc, $Z$ in kpc, period in days, Catalina amplitude and time-averaged magnitude $V_{\rm CSS}$, radial velocity, its uncertainty and the corresponding source, the $V_{\rm r}$ source, a quality and a corresponding comment, and finally the 2MASS and CRTS names when available. The first three lines of the catalogue are given in Table~3 in an abridged version for clarity.

\begin{table*}[!t]
\caption[]{Abridged version of the catalogue of M halo northern long period variables.}
\begin{center}
\begin{tabular}{lrrrrrrrrrrrrr}
\noalign{\smallskip}
\hline
\hline
\noalign{\smallskip}
 Star & $\alpha $~~~ & $\delta$~~~& $E_{\rm B-V}$  &  $(\KS)_0$ & $(J-\KS)_0$& $M_{\rm Ks}$ & $D$ & $\delta D$ & $P$ & 
 $\Delta V$ & $V_{\rm CSS}$ & $V_{\rm r}$ & $\delta V_{\rm r}$ \\
\noalign{\smallskip}
\hline
\noalign{\smallskip}

MMK-001 &   4.06675&  $-$17.60342&  0.039&  10.39&  1.10 &  $-$6.44 &  23.2 & 4.0 &   134.0 &   1.5 & 13.83 &   $-$59.42 &  6.8\\
MMK-002 &   5.64850&   $-$5.20222&  0.026&   9.81&  1.07 &  $-$6.24 &  16.2 & 1.5 &   424.4 &   0.4 & 13.49 &  $-$116.00 &  0.3\\ 
MMK-003 &   8.55188&   $-$1.53182&  0.019&  14.15&  0.93 &  $-$6.62 & 142.2 &24.5 &   138.2 &   1.5 & 17.01 & $-$9999.00 &  0.0\\ 

\noalign{\smallskip}
\hline
\noalign{\smallskip}
\end{tabular}
\end{center}
\end{table*}

%==============================================================================================================

\section{Conclusions}

This work can be summarized as follows:

{\noindent} (1) By considering available literature on long period variables out of the galactic plane and at $\delta > -20$\degree,  our goal was to build a catalogue of AGB M stars located in the halo. This catalogue will make a northern complement to the  census of C stars established by Huxor \& Grebel (2015). After merging and cleaning  available catalogues, a sample of 417 pulsating stars in the northern hemisphere is achieved, with no known C stars included.  Of these 417 stars, 262  are located at more than $\sim$\,5\,kpc  from the galactic plane. Consideration of their $(J-K_{\rm s})_0$ colour, periods, and amplitudes suggests that they are M stars. 

{\noindent} (2) By selecting  stars far away from the galactic plane ($|Z| >$\,5~kpc), one confirms a strong concentration along the apparent, orbital path of the Sgr tidal arms, especially on the leading arm. A major finding of this study is that there are about three times more M stars than C stars. This offers a future opportunity to investigate the Sgr AGB population  in more detail and analyse it in a complementary way to C stars.

{\noindent} (3) The pulsation modes, needed for distance determination,  are  found with a diagram showing  the $(I-K_{\rm s})_{\rm 0}$ colour as a function of   period, and by comparing it to that of the Large Magellanic Cloud. As for C stars, very few M stars are found to be on sequence C'.  Then, near-infrared Wesenheit P-L relations established for LMC oxygen-rich AGB stars are adopted. Distances up to $\sim$ 150\,kpc are derived, with typical relative uncertainties better than 20\%. Two pulsating stars found in the Pal~4 globular cluster have correct distances of 100~kpc. 

{\noindent} (4) If one considers stars located within 13\degree\, from the Sgr mean orbit plane, our distances for the leading arm agree very well with those recently derived with RR Lyr stars. Some of them are somewhat larger in the trailing arm. We find four stars located at the apocenter. Two stars are found to be close  to two carbon stars in the A16 substructure at $\sim$\,150\,kpc. There are seven stars angulary close to the Sgr orbit plane and more distant  than 70 kpc. We also draw attention to a clump of M and C stars in the halo angularly close to the Orphan stream, but distinct of it.  Radial velocities are required to more rigorously check these conclusions in these cases.

{\noindent} (5)  A catalogue of M AGB stars is presented. These stars belong to the thick disk, the halo, and the Sgr arms. Ninety radial velocities were collected from $Gaia$ and other surveys like RAVE. Significant proper motions were used to check supposedly large distances, but few  are discarded and flagged in the catalogue. Gaia DR2 parallaxes are presently too uncertain to achieve a robust comparison with our distances. The catalogue will be available on Vizier at CDS, or on request. Further investigation of our sample with additional radial velocities and Gaia variability information will be presented in a future paper.

\appendix
\section{}

The goal of this appendix is to describe how we estimate the $I_0$-band time-averaged photometry of our sample, which is needed to separate sequence C and C'.   From our sample of 417 stars, we identify  a subsample of 96 stars with  good  quality $I$-band photometry from the DENIS catalogue (uncertainty on $I$  $<$ 0.2 mag, 90\% of them with $<$ 0.05 mag error). Then, the Catalina $V_{\rm CSS}$, the 2MASS $J$ \& $K_{\rm s}$, and the DENIS $I$ are corrected for interstellar extinction by adopting the color excess from Schlegel et al.\ (1998). The used extinction law is from Cardelli et al.\ (1989), that is, $A_{\rm V} = 3.107 E(B-V)$, $A_{\rm I} = 1.705 E(B-V)$, $A_{\rm J} = 0.804 E(B-V)$, $A_{\rm K} = 0.342 E(B-V)$. Finally, we search for a function $f$($(V_{\rm CSS})_0$, $J_0$, $(K_{\rm s})_0$) that is as close as possible to the observed $I_0$. We find that the best fit relation is:\\

\noindent $I_0 =J_0 -0.22 + 0.58$\,$ ((V_{\rm CSS})_0 - J_0) + 0.05$\,$( (V_{\rm CSS})_0 -13.5)^{2}$ \\

This fit  is found with $N=96$ stars, with a scatter $\sigma$\,$=$\,$0.22$~mag. Outliers at more than 2.2 $\sigma$ were removed. We checked that the residuals do not depend on color such as $(V_{\rm CSS}-(K_{\rm s})_0)$. These 96 stars have the following properties  with $ 10.5 < (V_{\rm CSS})_0 < 17$, $ 2.6 < (V - K_{\rm s})_0 < 5.6$, $ 0.75 <(J-K_{\rm s}) < 1.4$, $ E(B-V) < 0.3$. 
These properties are obeyed by 80\% of stars in our sample.

%%%%%%%%%%%%%%%%%%%%%%%%%%%%%%%%%%%%%%%%%%      
\begin{acknowledgements}
N.M.  thanks our anonymous referee for remarks that greatly improved the mansucript. We also thank Olivier Richard for generous help, Henri Reboul and  Denis Puy for encouragments, and the staff of LUPM for kind assistance.  We appreciated remarks from Patrick de Laverny  on a preliminary version of the manuscript. N.M. is grateful to the Universit\'e de Montpellier for support. This work makes use of the Catalina database (California Institute of Technology, NASA), the Lincoln Near-Earth Asteroid Research LINEAR database (Massachusetts Institute of Technology Lincoln Laboratory, NASA and US Air Force), the Two Micron All Sky Survey (University of Massachusetts and IPAC/Caltech funded by NASA and NSF), the European infrared photometry {\sc DENIS} experiment. 
We exploited the data of the European Space Agency mission {\it Gaia} processed by the {\it Gaia} Data Processing and Analysis Consortium, and the {\sc SIMBAD}/Vizier facilities, offered by and operated at CDS, Strasbourg, France.
\end{acknowledgements}
%-------------------BIBLIOGRAPHY---------------------------------------------------

\end{document}